\documentclass[12pt,a4paper]{panl}
\usepackage{cite}
\usepackage{wrapfig}
\usepackage{graphicx}
\usepackage{amssymb}
\usepackage{amsfonts}
\usepackage{amsmath}
\usepackage{longtable}
\usepackage{rotating}
\usepackage{lscape}
\usepackage{epsfig}
\usepackage{multirow}
\usepackage{subfigure}
\usepackage{lineno}

\originalTeX
\begin{document}

\title{Proton and carbon-ion minibeam therapy: from modeling to treatment}
\maketitle
\authors{I.A.\,Pshenichnov$^{a}$\footnote{E-mail: pshenich@inr.ru}, U.A.\,Dmitrieva$^{a}$, S.D.\,Savenkov$^{a}$, A.O.\, Svetlichnyi$^{a}$}
\setcounter{footnote}{0}
\from{$^{a}$\,Institute for Nuclear Research, Russian Academy of Sciences, Moscow, 117312 Russia }

\begin{abstract}
Arrays of minibeams of protons and $^{12}$C in tissue-like media were modeled with Geant4 toolkit. A set of beam energies was used in simulations to provide a Spead-out Bragg peak (SOBP) extended by 6~cm in depth for protons as well as for $^{12}$C. In both cases, beams of 0.3~mm or 0.5~mm FWHM were arranged at the entrance to a water phantom either on a rectangular or an hexagonal grid to compare two kinds of projectiles and different minibeam patterns.  Differential and cumulative dose-volume histograms (DVH) were calculated and compared for protons and $^{12}$C as dose uniformity metrics. A uniform dose distribution was easily achieved with protons due to an enhanced lateral scattering of these projectiles in comparison to $^{12}$C. The cumulative DVHs calculated for 0.3~mm or 0.5~mm minibeams almost coincide in the target volume, but diverge for different grid patterns. In contrast, cumulative entry DVHs were found similar for both grid patterns, but different for 0.3~mm and 0.5~mm minibeams. 

\end{abstract}
\vspace*{6pt}

\noindent
PACS: 87.55.Gh, 87.55.K$-$


\label{sec:Intro}
\section*{Introduction}

To date, more than {360 000} patients have been treated at more than 100 proton and 8 heavy-ion cancer therapy facilities around the world~\cite{PTCOG2022}. While beams of protons, $^{4}$He and $^{12}$C nuclei are very effective in killing deep-seated solid tumors, some damage to surrounding healthy tissues is  unavoidable and must be reduced. Multiple treatment fields from different directions are commonly used to redistribute the entry dose and alleviate the damage outside the tumor volume. Alternatively, arrays of thin parallel proton beams, spatially fractionated at the entrance to the patient's body, were proposed for treatment~\cite{Prezado2013},\cite{Zlobinskaya2013}. As shown in a preclinical study~\cite{Lamirault2020}, proton minibeams spare normal tissues on the way to the target volume where minibeams become wider and finally overlap. 
Since  many normal cells between individual minibeams receive a minimal dose, this facilitates the recovery of normal tissues compared to the traditional particle therapy characterized by uniform entrance dose. 
As demonstrated~\cite{Prezado2018b}, proton minibeam therapy widens the therapeutic index for high-grade gliomas. It can be expected that a reduced damage to healthy tissues from minibeams will allow to reduce the number of treatment fields, thereby reducing the time spent on the entire sequence of patient positioning and treatment planning procedures. This will ultimately reduce the overall cost of treatment.

The implementation of proton or heavy-ion minibeam radiation therapy (MBRT) is technically challenging in terms of precise shaping of sub-millimeter beams with grid collimators~\cite{Lamirault2020} or by magnetic focusing~\cite{Datzmann2020}. At the same time, detailed modeling of mini-beam propagation in tissue-like media is needed to plan future {\em in-vivo} studies to link the peak-to-valley dose ratio (PVDR)~\cite{Prezado2013}, therapeutic advantage (TA)~\cite{Charyyev2020} or other metrics with the effectiveness of the minibeam therapy. In the present work 3D dose distributions in water from arrays of 0.3/0.5~mm FWHM parallel Spead-out Bragg peak (SOBP) proton and $^{12}$C beams were calculated for rectangular and hexagonal minibeam patterns. The dose uniformity in the target tumor volume was evaluated by calculating differential and cumulative dose-volume histograms (DVH). DVHs were compared for protons and $^{12}$C at the entry to the phantom as well as in the target volume. A deep-seated target volume (6~cm wide, centered at 13~cm depth) to be irradiated with energetic protons and $^{12}$C was considered because the sparing of a large volume of normal tissues is crucial in this case. While rectangular and hexagonal minibeam patterns have been compared for protons~\cite{Sammer2017}, a similar study for $^{12}$C is missing, and the present work aims to fill this gap.

\label{sec:Methods}
\section*{Materials and methods}

Arrays of minibeams of protons and $^{12}$C propagating in a water phantom of $10\times10\times200$~mm$^3$ size were modelled with Geant4 toolkit~\cite{Allison2016} of version 10.3. The physics lists used in modeling were the same as in our previous work~\cite{Dewey2017}, namely: electromagnetic processes were modeled with {Standard\_opt3} list, Binary Cascade (BIC) model was involved for proton- and neutron-induced nuclear reactions and Quantum Molecular Dynamics (QMD) model for nucleus-nucleus collisions. 

\begin{figure}[!htb]
\begin{centering}
\begin{minipage}{0.8\columnwidth}
\includegraphics[width=0.5\columnwidth]{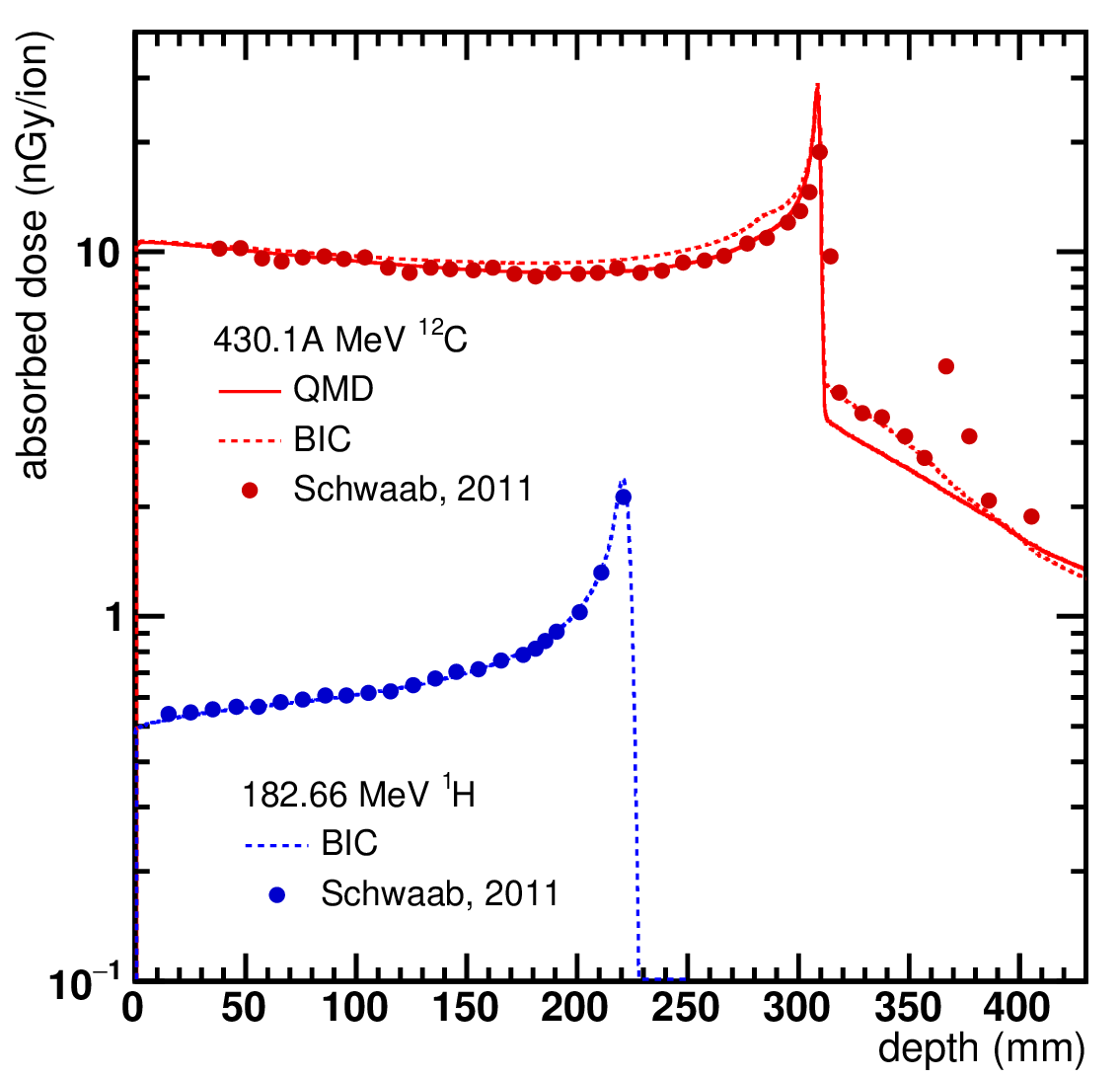}
\includegraphics[width=0.5\columnwidth]{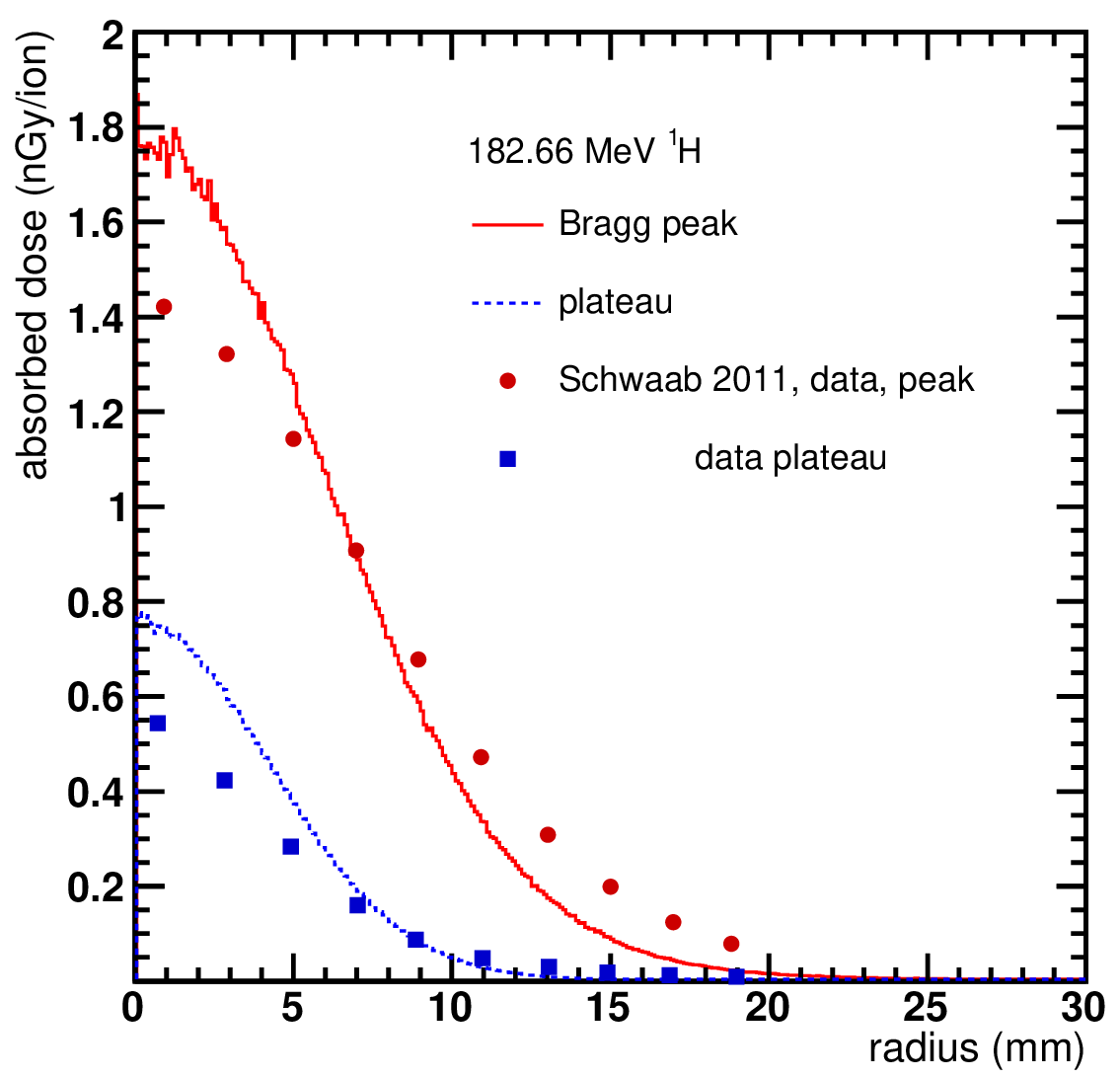}
\end{minipage}
\caption{Left: depth-dose distributions for 182.66~MeV proton and 430.1A~MeV $^{12}$C pencil-like beams in water calculated with QMD model (solid-line histograms) and BIC model (dashed-line histogram). Right: lateral dose distributions for 182.66~MeV proton pencil-like beams at the plateau (dashed-line histogram) and Bragg peak (solid-line histogram). Points represent measurements~\cite{Schwaab2011}.}
\label{fig:depth-dose}
\end{centering}
\end{figure}

Prior to conducting modeling of arrays SOBP minibeams, our Geant4-based code was validated with data on depth-dose and lateral dose distributions measured for monoenergetic pencil-like beams~\cite{Schwaab2011}, as shown in Fig.~\ref{fig:depth-dose}. All dose distributions were calculated per beam particle and the respective experimental data were rescaled for comparison with calculations. One can note a good description of measured depth-dose distributions for protons and $^{12}$C  by calculations, respectively, with BIC and QMD models. However, the lateral spread of proton pencil-like beams is partly underestimated in modeling.  Wider transverse profiles of pencil-like beams calculated with Geant4 versions~9.2 and 10.2 in comparison to measurements and other transport models were also reported, respectively, in Refs.~\cite{Grevillot2010} and~\cite{Solie2017}. One can note that such comparisons of measured and calculated distributions were performed for pencil-like beams of few millimeter FWHM. Similar validation for minibeams with $\sim 0.1$~mm FWHM will be a subject of future studies, since respective data are not yet available.

Two sets of beam energies and their weights optimized, respectively, for protons and $^{12}$C to provide flat SOBP distributions of biological dose~\cite{Dewey2017} were used in the present work. The calculations of the relative biological effectiveness (RBE) were performed on the basis of the modified microdosimetric kinetic model (MKM)~\cite{Kase2006}. The biological endpoint for the SOBP optimization was chosen as 10\% survival (at 5.1~Gy(RBE)) of human salivary gland (HSG) tumor cells ($(\alpha/\beta)_{\mathrm{x-rays}} = 3.8$~Gy) in the target volume.  The resulting SOBPs of 6~cm width were centered at 13~cm depth.

\label{sec:Minibeams}
\section*{Modelling minibeams in water}

\begin{figure}[!htb]
  \centering
  \subfigure{\includegraphics[width=0.75\columnwidth]{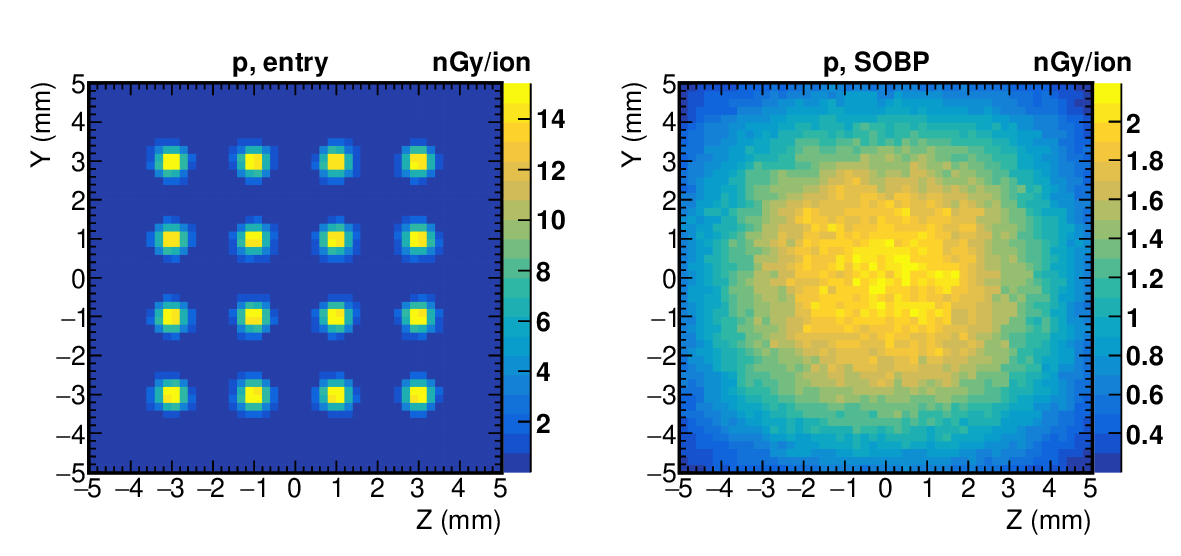}} 
  \subfigure{\includegraphics[width=0.75\columnwidth]{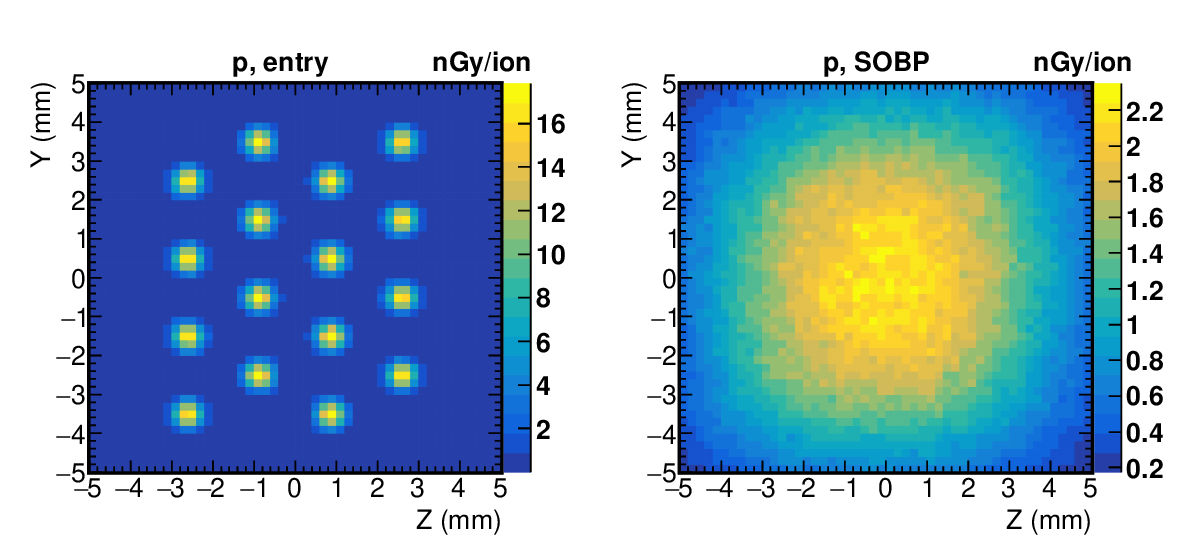}} 
\caption{Rectangular (top) and hexagonal (bottom) patterns of minibeams of protons of 0.5~mm FWHM in water. Dose distributions were calculated at the entry to phantom at 0--20~mm depth (left) and at the target volume at 100--160~mm depth (right).} 
  \label{fig:H1_mini}
\end{figure}

\begin{figure}[!htb]
  \centering
  \subfigure{\includegraphics[width=0.75\columnwidth]{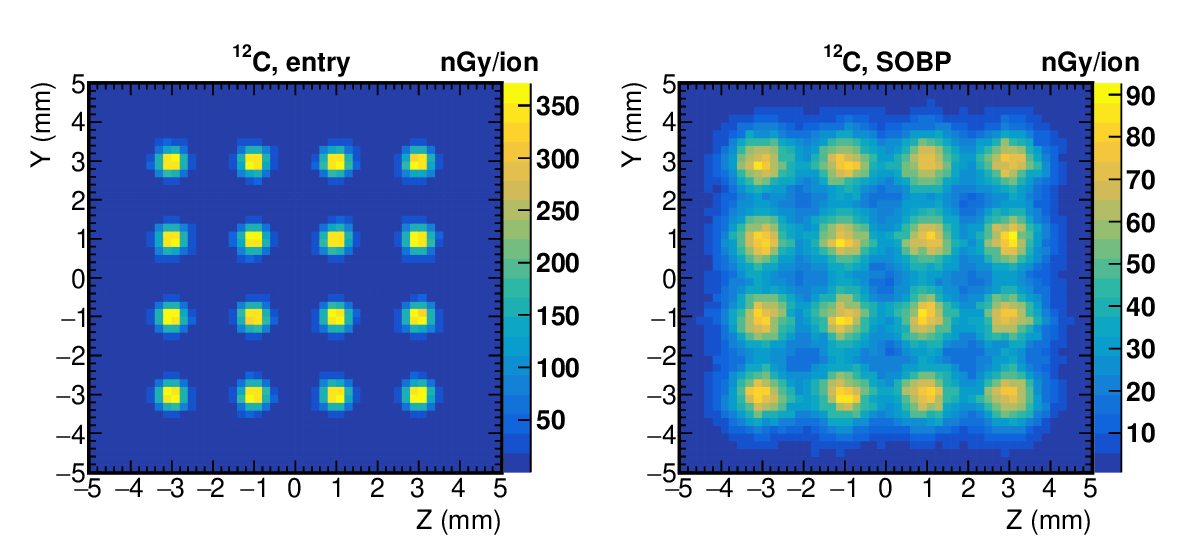}} 
  \subfigure{\includegraphics[width=0.75\columnwidth]{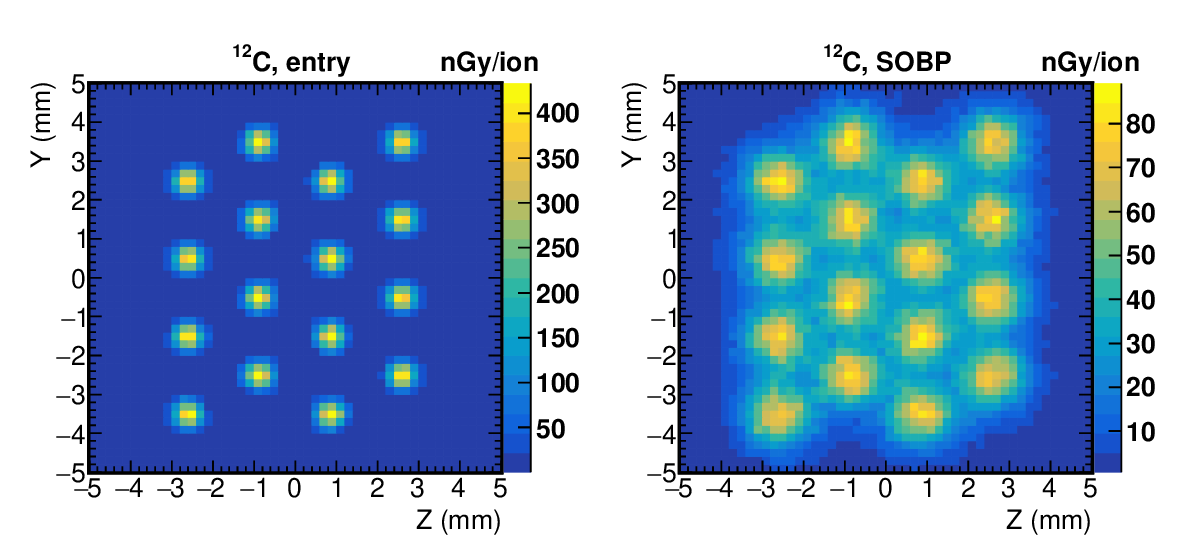}} 
  \caption{Same as in Fig.~\protect\ref{fig:H1_mini}, but for $^{12}$C minibeams in water.} 
  \label{fig:C12_mini}
\end{figure}

Rectangular and hexagonal patterns of minibeams of protons and $^{12}$C, with the distance between beam centers of 2~mm, were used in modeling as presented in Figs.~\ref{fig:H1_mini} and~\ref{fig:C12_mini} for beams of 0.5~mm FWHM. Similar modeling was performed for minibeams of 0.3~mm FWHM  also assuming Gaussian intensity profile of each minibeam with zero angular divergence at the entry surface of the phantom. Transverse dose distributions at the entrance to phantom were calculated for the average dose at 0--20~mm depth. As mentioned above, sets of projectile energies were used to provide the flat SOBP distributions of RBE-weighted dose from 100~mm to 160~mm in depth in the target volume. Therefore, the same depth interval of 100--160~mm was used to calculate the average transverse distributions in the target volume shown in Figs.~\ref{fig:H1_mini} and~\ref{fig:C12_mini}. 

Initially thin minibeams become wider with depth and partially overlap with each other due to the multiple Coulomb scattering of protons and $^{12}$C during their propagation in water. The production of secondary nuclei in fragmentation of $^{12}$C also takes place, see Ref.~\cite{Dewey2017}.  Since the lateral spread of dose from protons in comparison to $^{12}$C is much larger, a uniform dose distribution is delivered by protons to the target volume, see Fig.~\ref{fig:H1_mini}. However, for the considered arrangement of $^{12}$C minibeams, peak and valley doses from $^{12}$C remain distinct in the target SOBP volume. One can propose a shorter center-to-center distance for irradiation with $^{12}$C minibeams to achieve a better dose uniformity in the target volume. In order to make a quantitative assessment of the dose uniformity, e.g., in the target volume, a certain dose distribution metrics should be used.  

\label{sec:DVH}
\section*{Dose-volume histograms for protons and $^{12}$C}

Dose distributions from minibeams are usually characterized by their PVDR, as in Ref.~\cite{Prezado2013}. To obtain this metrics, the ratio between the maximum and minimum local dose is calculated, in particular, for arrays of planar minibeams. PVDR can be also safely defined for the hexagonal minibeam grid, where the extremum values are unambiguously identified, respectively, at the beam centers and in the middle of the distance between them, see Fig.~\ref{fig:C12_mini}. However, as also seen from Fig.~\ref{fig:C12_mini}, the rectangular minibeam grid reveals two distinct local minima: (1) on the sides of the square cell, and (2) on its diagonal. Instead of PVDR, differential and cumulative dose-volume histograms (DVH) can be considered for rectangular and hexagonal minibeam patterns. DVH are widely used in treatment planning in traditional radiation therapy~\cite{Drzymala1991}.

\begin{figure}[!htb]
\begin{centering}
\begin{minipage}{0.85\columnwidth}
\includegraphics[width=0.5\columnwidth]{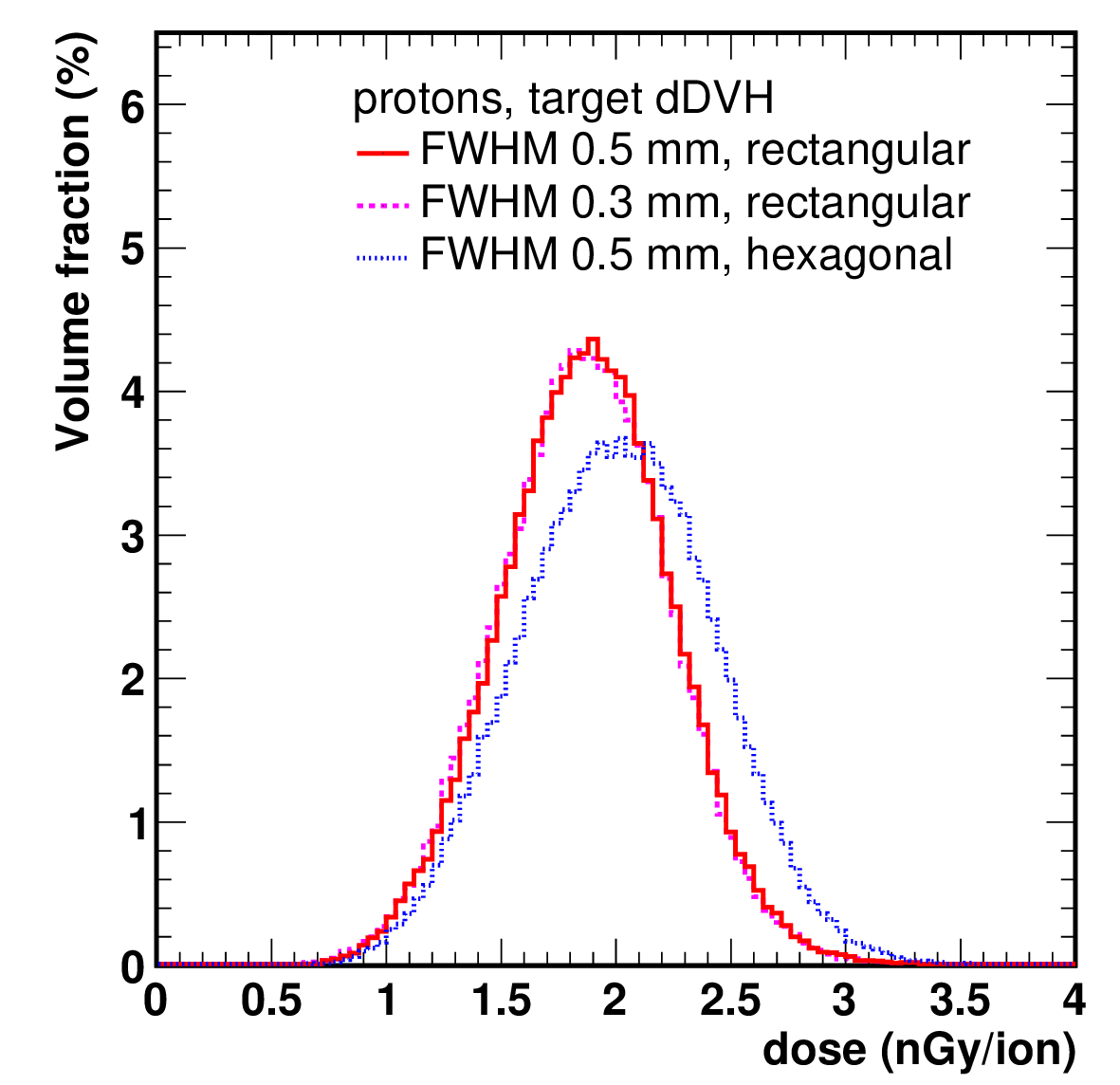}
\includegraphics[width=0.5\columnwidth]{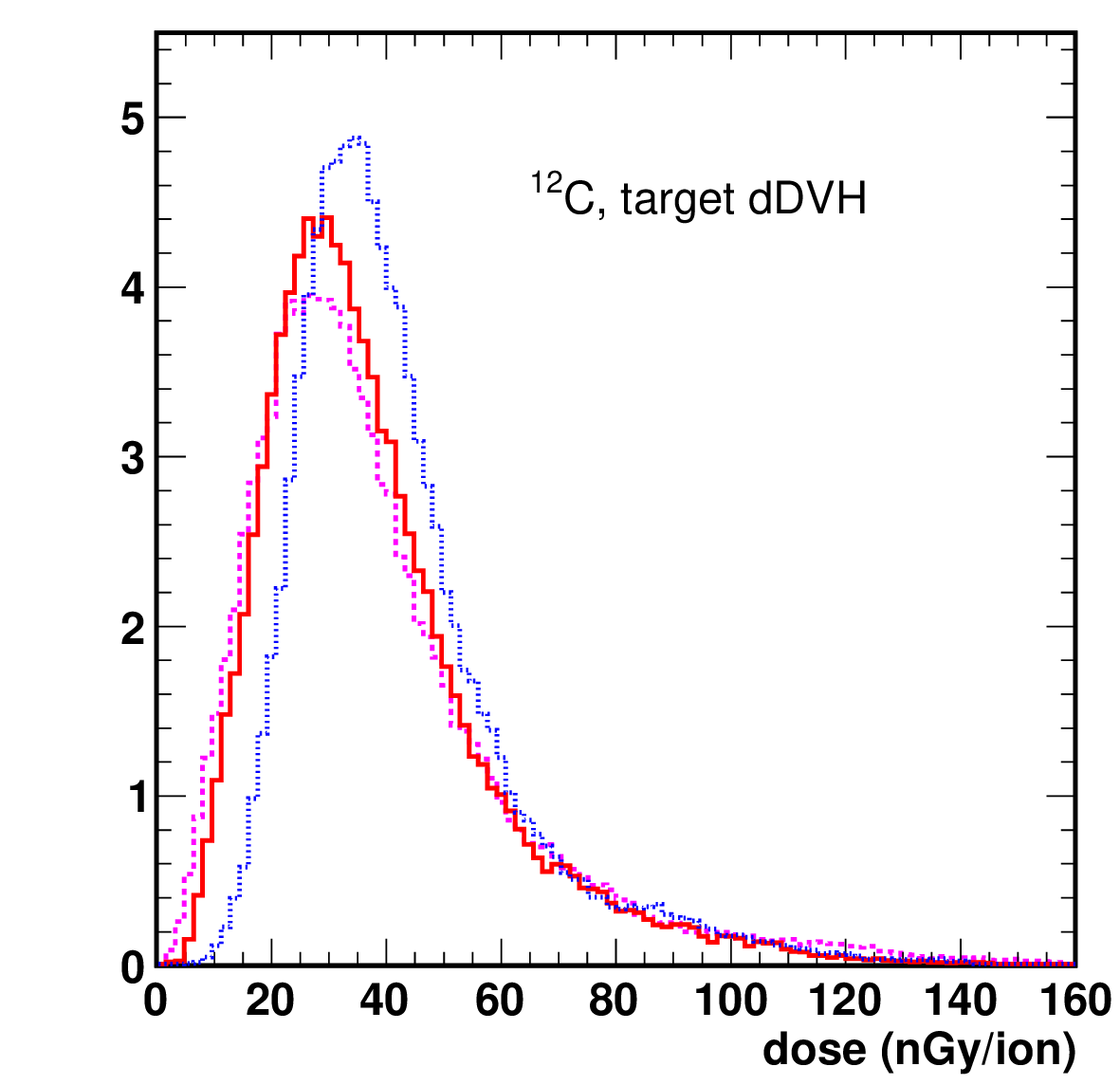}
\end{minipage}
\caption{Differential dose-volume histograms calculated in the target volume for rectangular and hexagonal patterns of proton (left) and $^{12}$C (right) minibeams of 0.3~mm and 0.5~mm FWHM.}
\label{fig:DVH_target}
\end{centering}
\end{figure}

Differential DVHs representing a fraction of target volume which receives a certain dose are shown in Fig.~\ref{fig:DVH_target} for rectangular and hexagonal patterns of proton and $^{12}$C minibeams of 0.3~mm and 0.5~mm FWHM. A central subvolume of $4\times4\times60$~mm$^3$ size was used in calculating target DVHs to avoid a dose reduction on the phantom boundary. Naturally, the results obtained for this subvolume as an unit cell can be later extended to much wider therapeutic fields with the same arrangement of minibeams on the basis of periodic boundary conditions. As seen from Fig.~\ref{fig:DVH_target}, differential DVHs calculated for 0.3~mm and 0.5~mm FWHM are very similar for proton as well as for $^{12}$C minibeams. However, minbeams of 0.5~mm, but arranged on the hexagonal grid demonstrate slightly higher average doses compared to the rectangular grid. This is explained, in particular, by a more dense arrangement of beams on the hexagonal grid. A long tail which extends up to 140~nGy/ion, well above the median dose of $\sim 30$~nGy/ion, is seen for $^{12}$C DVH, while low-dose voxels with $< 20$~nGy/ion are also present. All this indicates a high non-uniformity of the dose delivered to the target volume by $^{12}$C. In contrast, the dose delivered to the target by proton minibeams is strictly within the interval of 1--3 nGy/ion, without nonirradiated voxels. It should be reminded that a typical variation of the dose within the SOBP width for each individual minibeam is  $\pm 20$\% from its average SOBP dose to provide a uniform RBE-weighted dose in the target volume. Therefore, the dispersion seen in Fig.~\ref{fig:DVH_target} also includes this variation.       

\begin{figure}[!htb]
    \centering
    \includegraphics[width=0.75\columnwidth]{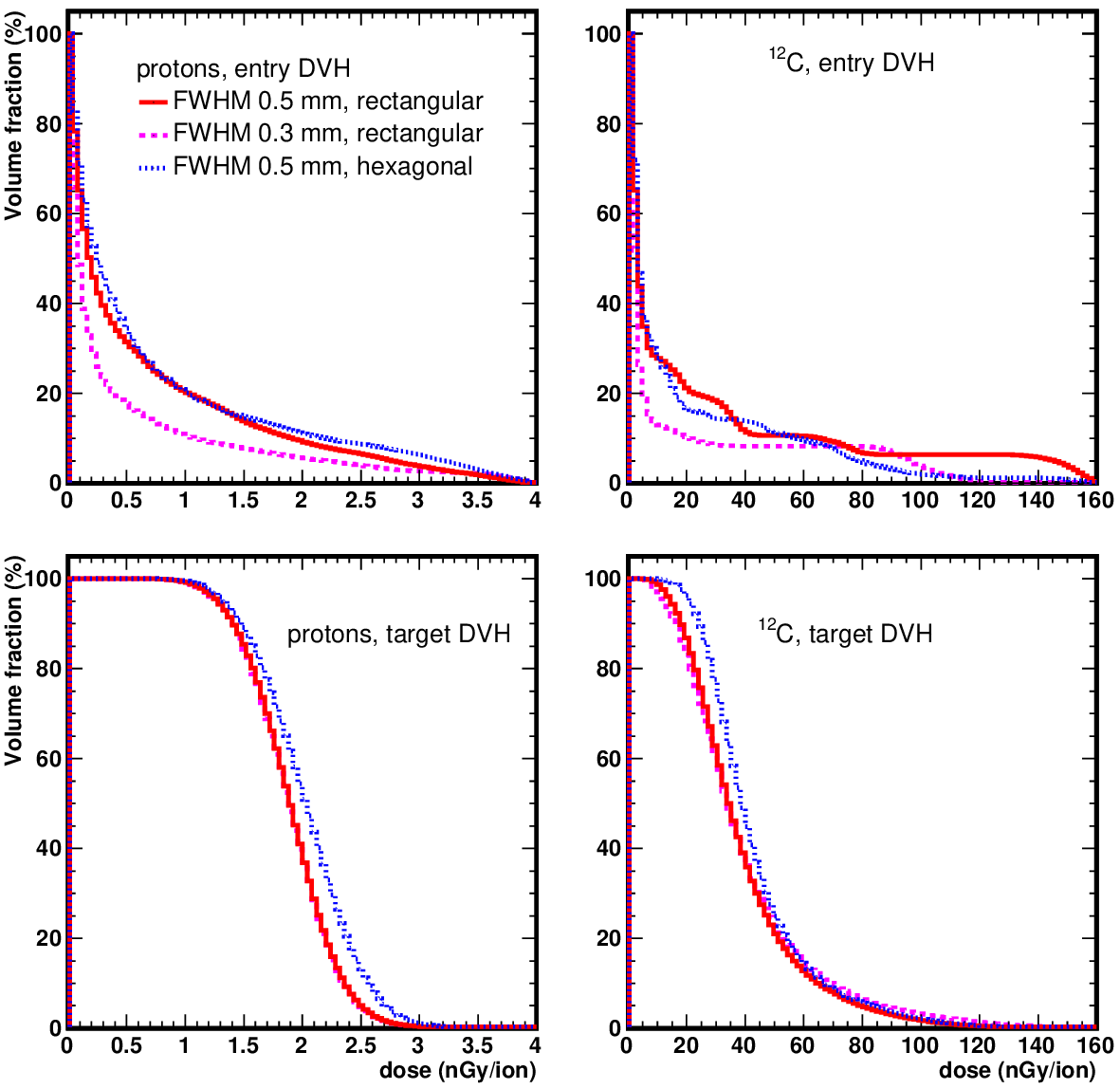}
    \caption{Cumulative dose-volume histograms for proton (left) and $^{12}$C (right) minibeams calculated at the entrance to the phantom (top) and in the target volume (bottom).}
    \label{fig:cDVH}
\end{figure}

Cumulative DVHs (or simply DVHs) for proton and $^{12}$C minibeams calculated at the entrance to the phantom and in the target volume are presented in Fig.~\ref{fig:cDVH}. By definition~\cite{Drzymala1991}, each DVH provides the fraction of cells (or voxels) (in \%) receiving greater or equal of a given dose represented by values on its x-axis. For example, from the DVH representing the dose distribution from protons (Fig.~\ref{fig:cDVH}), one can immediately see that all voxels in the target volume receive at least 1~nGy/ion, and about half of all voxels receive above 2~nGy/ion. Entry DVHs also presented in Fig.~\ref{fig:cDVH} demonstrate sensitivity to FWHM of proton and $^{12}$C minibeams. As seen, the DVH calculated with thinner beams of 0.3~mm FWHM are bent closer to (0,0), thus providing more favorable conditions to spare healthy tissues. However, beams of 0.5 FWHM provide similar entry DVHs for both minibeam patterns. This means that different techniques~\cite{Lamirault2020},\cite{Datzmann2020}, with different minibeam patterns, can be equally effective in sparing normal tissues outside the tumor volume as far as the FWHM and center-to-center distance are equal.   

\label{sec:conclusion}
\section*{Conclusion}

Several configurations of arrays of minibeams of protons and $^{12}$C in water were modeled with Geant4 of version 10.3. Rectangular and heaxagonal patterns of minibeams with the center-to-center distance of 2~mm were simulated, and in both cases 6~cm-wide SOBP peaks were centered at 13~cm depth to assess the ability of protons and $^{12}$C to provide dose uniformity evaluated in terms of differential and cumulative dose-volume histograms.  It was found that because of their enhanced lateral scattering proton minibeams provide much more uniform dose distributions in the target volume in comparison to $^{12}$C minibeams. DVHs calculated for protons and $^{12}$C in the target volume for 0.3~mm and 0.5~mm minibeams almost coincide, but some differences are seen between rectangular and hexagonal patterns. In contrast, the arrays 0.3~mm and 0.5~mm minibeams are found different in their effectiveness to spare tissues outside the target volume according to the DVHs calculated at the entrance to the phantom. However, this effectiveness is less sensitive to the minibeam pattern geometry at the entry to the phantom. The present study provides a basis for detailed calculations of RBE and other radiobiological properties of minibeams for planning future pre-clinical in-vivo studies.

We are grateful to the Russian Science Foundation for the support within the grant No 23-25-00285 “Modelling of the physical and biological properties of therapeutic minibeams of protons and nuclei”.

\bibliographystyle{pepan}
\bibliography{Pshenichnov_Minibeams_Modelling}

\end{document}